\documentclass[lettersize,journal]{IEEEtran}
\usepackage{stmaryrd}

\usepackage{graphicx}
\usepackage{amsmath,amsfonts,amssymb}
\usepackage{algorithmic}
\usepackage{algorithm}
\usepackage{array}
\usepackage[caption=false,font=normalsize,labelfont=sf,textfont=sf]{subfig}
\usepackage{textcomp}
\usepackage{stfloats}
\usepackage{url}
\usepackage{verbatim}
\usepackage{cite}
\usepackage{bm}
\usepackage{booktabs}
\usepackage{multirow}
\usepackage{hyperref}
\usepackage{tikz}
\usetikzlibrary{patterns,shapes,arrows,arrows.meta,chains,fit,positioning,calc,backgrounds}
\usepackage{pgfplots}
\pgfplotsset{compat=newest}

\graphicspath{{images/}}

\DeclareMathOperator*{\med}{\text{med}}

\newcommand{\fig}{Figure~}
\newcommand{\tab}{Table~}

\newcommand{\rotangle}{\alpha}
\newcommand{\layeroverlap}{\tau}
\newcommand{\interpolationkernel}{\zeta}
\newcommand{\lossalpha}{\theta}

\newcommand{\channel}{v}
\newcommand{\centercam}{c}

\newcommand{\calcoordM}{m}
\newcommand{\calcoordN}{n}
\newcommand{\calcoord}{\calcoordM_\channel, \calcoordN_\channel}
\newcommand{\descoordX}{x}
\newcommand{\descoordY}{y}
\newcommand{\descoord}{\descoordX, \descoordY}

\newcommand{\addmask}{\kappa}

\newcommand{\scalex}{s_x}
\newcommand{\scaley}{s_y}
\newcommand{\scaler}{s_r}

\newcommand{\cameras}{E}

\newcommand{\disparityoperator}{d}

\newcommand{\homograhpy}{\bm{H}}

\newcommand{\lowLRA}{A^l}
\newcommand{\lowLRB}{B^l}
\newcommand{\highLRA}{A^h}
\newcommand{\highLRB}{B^h}
\newcommand{\disparity}{D}
\newcommand{\calibrated}{C}
\newcommand{\msmasked}{K}
\newcommand{\mask}{O}
\newcommand{\msimage}{M}

\newcommand{\disparitylayer}{L}
\newcommand{\networkoutput}{N}

\newcommand{\disparitywarped}{W}

\begin{document}

\title{Multispectral Snapshot Image Registration Using Learned Cross Spectral Disparity Estimation and a Deep Guided Occlusion Reconstruction Network}

\author{Frank Sippel,~\IEEEmembership{Graduate Student Member,~IEEE}, Jürgen Seiler,~\IEEEmembership{Senior Member,~IEEE}, and André Kaup~\IEEEmembership{Fellow,~IEEE}
\thanks{Manuscript received March XX, 2024; revised June XX, 2024.}
\thanks{\textit{Corresponding author: Frank Sippel}}
\thanks{The authors gratefully acknowledge that this work has been supported by the Deutsche Forschungsgemeinschaft (DFG, German Research Foundation) under project number 491814627.}
\thanks{The authors are with the Chair of Multimedia Communications and Signal Processing, Friedrich-Alexander-Universität Erlangen-Nürnberg (FAU), Erlangen, Germany (e-mail: frank.sippel@fau.de; juergen.seiler@fau.de; andre.kaup@fau.de).}
}

\markboth{Multispectral Snapshot Image Registration,~Vol.~XX, No.~X, June~2024}%
{Sippel \MakeLowercase{\textit{et al.}}: Multispectral Snapshot Image Registration Using Learned Cross Spectral Disparity Estimation and a Deep Guided Occlusion Reconstruction Network}

\IEEEpubid{0000--0000/00\$00.00~\copyright~2024 IEEE}

\maketitle

\begin{abstract}
Multispectral imaging aims at recording images in different spectral bands.
This is extremely beneficial in diverse discrimination applications, for example in agriculture, recycling or healthcare.
One approach for snapshot multispectral imaging, which is capable of recording multispectral videos, is by using camera arrays, where each camera records a different spectral band.
Since the cameras are at different spatial positions, a registration procedure is necessary to map every camera to the same view.
In this paper, we present a multispectral snapshot image registration with three novel components.
First, a cross spectral disparity estimation network is introduced, which is trained on a popular stereo database using pseudo spectral data augmentation.
Subsequently, this disparity estimation is used to accurately detect occlusions by warping the disparity map in a layer-wise manner.
Finally, these detected occlusions are reconstructed by a learned deep guided neural network, which leverages the structure from other spectral components.
It is shown that each element of this registration process as well as the final result is superior to the current state of the art.
In terms of PSNR, our registration achieves an improvement of over 3 dB.
At the same time, the runtime is decreased by a factor of over 3 on a CPU.
Additionally, the registration is executable on a GPU, where the runtime can be decreased by a factor of 111.
The source code and the data is available at \url{https://github.com/FAU-LMS/MSIR}.
\end{abstract}


\section{Introduction}
\label{sec:introduction}
In comparison to classical RGB imaging, multispectral systems record more bands, possibly also in non-visible spectral areas.
Multispectral cameras typically capture in between six and 16 different application-dependent spectral bands.
This additional information, in comparison to RGB cameras, is very useful for many applications.
For example, these cameras are used in recycling to sort materials~\cite{moroni_pet_2015}, in medicine to determine the degree of burn of skins~\cite{sowa_classification_2006}, in forensics to find out the age of blood~\cite{edelman_hyperspectral_2013}, or in agriculture by discriminating between parts that need fertilizer and water and those that are healthy~\cite{lima_monitoring_2020}.

\begin{figure}
    \centering
    \input{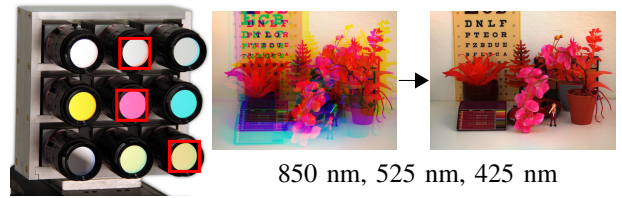}
    \vspace*{-0.3cm}
    \caption{The tackled problem of this paper. Images from a multispectral camera array (left) that are not usable when overlapped (middle) shall be registered (right).}
    \vspace*{-0.6cm}
    \label{fig:task}
\end{figure}

\begin{figure*}
    \centering
    \input{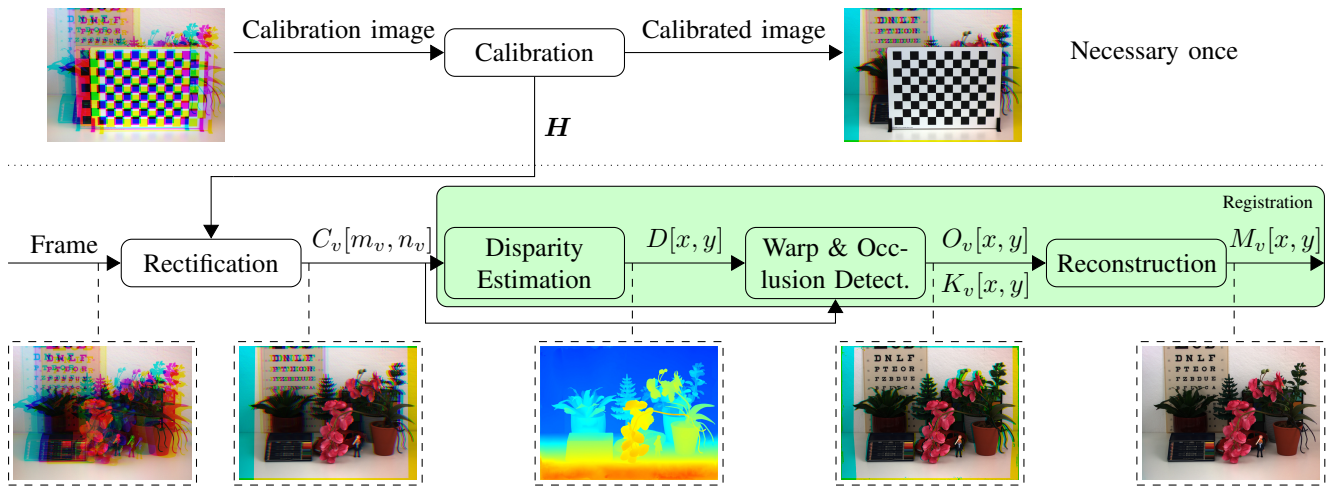}
    \caption{The whole pipeline necessary to calibrate and register multispectral images, including our proposed multispectral image registration. The calibration procedure might be necessary once for a video sequence, while the remaining elements are executed on all frames of a video individually.}
    \label{fig:pipeline}
\end{figure*}

Multispectral data cubes with two spatial dimensions and one spectral dimension, thus three-dimensional, can be captured in different ways.
One has to differentiate between scanning approaches and snapshot approaches.
Scanning solutions unfold the spectral dimension of the three dimensional data cube by either using one spatial dimension of a grayscale sensor to capture the light spectrum by using a dispersive element~\cite{hoye_pushbroom_2015}, or by using the time dimension to record each channel, e.g., by using a filter wheel~\cite{brauers_filterwheel_2008}.
In contrast, snapshot approaches try to capture the three dimensional data cube at once, therewith being able to produce multispectral temporal videos.
The high temporal resolution of multispectral videos is essential for many applications, for example, when measuring human brain activity~\cite{villringer_brain_1997}.

One approach to record multispectral video is by using spectral beamsplitters, which split a single beam of light into its components using semi-transparent mirrors~\cite{matchett_beamsplitter_2007}.
Unfortunately, these mirrors are not perfectly transparent for the light that should pass through.
Therefore, the usable number of filters is typically around 4.
Another solution is by using a coded aperture in addition to a dispersive element, such that each pixel on the sensor receives a mixture of different bands of different pixels~\cite{wagadarikar_cassi_2008}.
This results in a compressive sensing problem and thus needs a reconstruction process~\cite{yuan_cassialgo_2016}.
Furthermore, multispectral filter arrays~\cite{laypray_msfa_2014} extend the concept of a Bayer pattern~\cite{bayer_1976} to record even more channels, e.g., in a three by three pattern.
Like demosaicing algorithms for Bayer pattern sensors, multispectral filter arrays also need a reconstruction process to yield a dense multispectral data cube~\cite{mizutani_demosaic_2014}.
A problem with multispectral filter arrays is the low resolution of pixels actually recorded by the sensor.

\IEEEpubidadjcol
A promising approach to capture a multispectral data cube by snapshot recording is a camera array~\cite{genser_camsi_2020},
where each camera records a different spectral band~\cite{cam_array_notch_2022}.
This approach can either be used for multispectral imaging~\cite{cam_array_wheat_2021} or even hyperspectral imaging~\cite{cam_array_art_2022}.
In \fig\ref{fig:task}, a multispectral camera array arranged in a three times three grid is depicted.
In this array, identical industrial cameras with a spatial resolution of $1600 \times 1200$ pixels and the same lenses with a focal length of $16$ mm are used.
The main advantage of this approach is that different filters can be chosen according to the spectrum of interest, which can be optimally selected dependent on the application~\cite{sippel_optimal_2022}.
This is not possible using multispectral filter arrays or coded aperture approaches.
Genser et al.~\cite{genser_camsi_2020} have shown that a camera array approach to multispectral imaging is able to outperform multispectral filter arrays significantly.

Unfortunately, when imaging a scene using the camera array approach, the overlapped image of the raw data is not usable since every camera records from a different spatial position.
Thus, a calibration and a registration procedure is required.
The calibration has to be performed once and is not dependent on the actual input frame of the recorded multispectral video.
Calibration methods are very well researched and widely available.
In contrast, the registration procedure, which compensates the depth-based offset, is dependent on the input signal and thus needs advanced image processing.
Hence, the image registration needs to be performed on every frame of a multispectral video.
In this paper, we focus on this registration part.

The calibration and registration pipeline is depicted in \fig\ref{fig:pipeline}.
After the input images are rectified, the first step is a cross spectral disparity estimation, which indicates the pixel distance between corresponding pixels of the center view and the peripheral view.
This disparity map is then used to warp the peripheral images to the center view.
Furthermore, an occlusion detection is required, since the center camera sees pixels that are not recorded by all peripheral cameras.
Hence, these pixels are reconstructed in the final step of the pipeline.

\section{Related Work}
\label{sec:sota}
Multi-modal image registration is a general research area for overlaying two images of different modalities from different views~\cite{arar_regis1_2020}.
In literature, this often only includes estimating a rigid and a non-rigid transform, e.g., finding a homography transformation and a deformable field\cite{deng_regis2_2023}.
However, this neglects occlusions occurring when imaging a scene using camera arrays as well as the underlying geometry of the array.
Furthermore, due to these occlusions, missing pixels need to be reconstructed.
The state-of-the-art registration method for this regime is described by Genser et al.~\cite{genser_camsi_2020}.
This paper does not only present a registration procedure, but also introduces a calibration procedure for multispectral camera arrays.
This calibration procedure will also be used here when it is necessary to do so first.
Besides, all single elements of the multispectral image registration have related work as well.

\subsection{Disparity Estimation}
Disparity estimation is a prominent research topic, thus a lot of work has been presented in this area.
First, disparity estimation methods assuming the same spectral content for both views are shortly reviewed.
Chang et al.~\cite{chang_psmnet_2018} introduced Pyramid Stereo Matching Network (PSMNet), which can be viewed as the base work for learned disparity estimation.
They employ a siamese network for feature extraction and use them to calculate a cost volume.
This cost volume is then refined using a 3D convolutional neural network followed by a disparity regression.
This whole pipeline is end-to-end trainable.
The Guided Aggregation Network (GANet) by Zhang et al.~\cite{zhang_ganet_2019} introduces semi-global and local guided aggregation steps to refine the cost volume.
This approach was inspired by the classical semi-global matching~\cite{hirschmuller_sgm_2008} but can be backpropagated.
Xu et al.~\cite{xu_acvnet_2022} presented Attention Concatenation Volume Network (ACVNet), which uses a network to produce attention on the cost volume.
Afterwards, this cost volume is again refined using 3D convolutional neural networks.
Finally, Iterative Geometry Encoding Volume (IGEV) by Xu et al.~\cite{xu_igev_2023} uses an iterative update step to refine an initial disparity image based on this geometry encoding volume and a network-filtered version of the left image.

All of these methods use RGB image pairs to extract disparity and can hence not directly be used in multispectral imaging.
In the context of this paper, the images fed into the disparity estimation network are multimodal, since they are captured in different spectral bands.
Therefore, Zhi et al.~\cite{zhi_material_2018} and Han et al.~\cite{han_disp_2023} introduce a way to estimate disparity across spectral bands by using RGB and infrared images.
Genser et al.~\cite{genser_deep_2020} achieved a more general cross spectral disparity estimator by introducing a simple spectral augmentation algorithm.
Finally, GANet~\cite{zhang_ganet_2019} was retrained using a slightly improved data augmentation and local normalized images~\cite{sippel_cade_2024}.

\subsection{Occlusion Detection}
Occlusion detection is another essential part for image registration, since the peripheral cameras do not see all pixels that the center camera sees.
The goal of the occlusion detection is to detect these pixels.
For that, Li et al.~\cite{li_symmnet_2018} introduced a U-Net-based network to detect occlusion from left and right images.
This network architecture is a successor of the FlowNet architecture also producing occlusion maps~\cite{dosovitskiy_flownet_2015}.
Li et al.~\cite{li_symmnet_2018} also provide an analysis of different network architectures, which showed that this structure, where occlusions maps for both views are predicted jointly, produces the best results.
Ilg et al.~\cite{ilg_occlusion_2018} estimated occlusion directly in conjunction with the disparity map.
Note that occlusion detection based on neural networks lacks reliability and interpretability, which is a key ingredient in the presented multispectral image registration pipeline.
Pixels that need to be occluded, but are not, are assumed to be perfectly valid by the reconstruction module in the end.
Thus, the reconstruction network builds models based on these pixels, which then lead to false reconstructions of occluded pixels.

\subsection{Image Reconstruction}
A final step in the registration is to reconstruct the missing pixels detected by the occlusion detection.
Inpainting methods, operating only on the spectral component with occluded pixels, could be used to fill in these missing pixels.
Getreuer et al.~\cite{getreuer_total_2012} introduced a minimization problem using total variation for inpainting.
Of course, neural-network-based methods have been introduced as well.
The foundation was laid by Liu et al.~\cite{liu_image_2018} which modified the classic convolution by paying attention to missing pixels and employed partial convolution.
Free-form image inpainting by Yu et al.~\cite{yu_free_2019} uses two stages to first estimate a coarse result, which is then refined by a second neural network.
Moreover, contextual attention is employed.
Nazeri et al.~\cite{nazeri_edgeconnect_2019} also use two stages, however, here the missing edges are estimated first.
Subsequently, these edges are exploited to reconstruct the whole image.
Finally, also transformers~\cite{li_mat_2022} were successfully employed on the image inpainting problem.
The goal of inpainting methods is to fill in missing pixels to produce nice looking images.

However, in the case of our paper more information is available.
Missing pixels are only occurring in peripheral views, thus the fully preserved center view can be used as cross spectral guide for the reconstruction problem.
For that, prior work includes algorithms based on linear regression of which the coefficients are either calculated using a local neighborhood~\cite{genser_spectral_2018, genser_join_2020} or globally using non-local filtering~\cite{sippel_spatio_2021}.

\begin{figure*}
    \centering
    \input{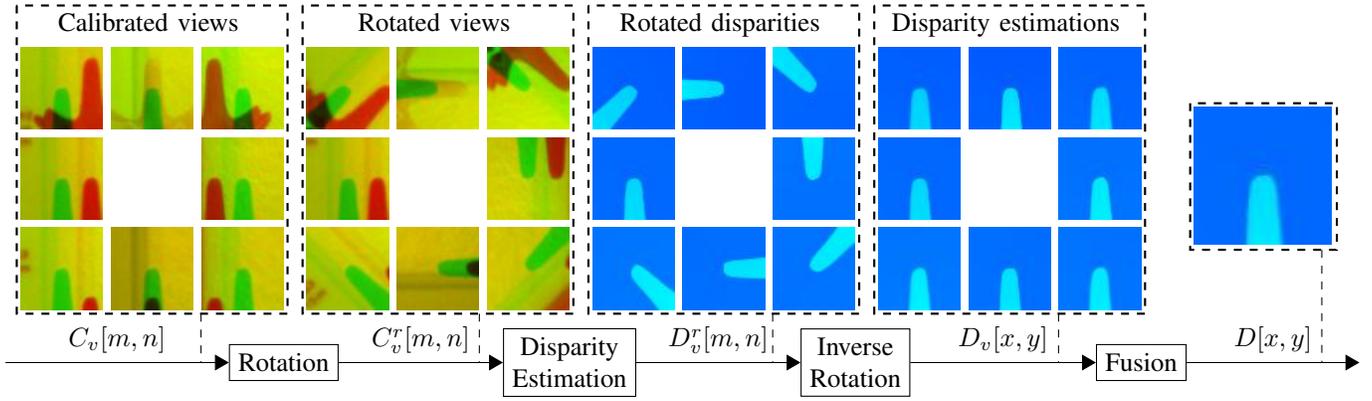}
    \caption{The process of estimating a single disparity map for the center view. The images on the left show false color images with center red channel being the center view and the green channel depicting the corresponding peripheral view. First the calibrated views are rotated such that the disparity is purely horizontal. The angle can be calculated by knowing the array geometry. Then, a cross spectral disparity estimator is executed on each pair individually. These estimates are rotated back and fused to a single estimate in the end.}
    \label{fig:disparity}
\end{figure*}

\section{Proposed Method}
\label{sec:method}
The goal of the proposed registration is to map every peripheral view of the array to the center view.
Therefore, each pixel should depict the exact same point of the same object in different spectral bands, which is then properly usable for example by classification and segmentation algorithms.
For that, the multispectral image registration procedure as shown in \fig\ref{fig:pipeline} is proposed to register the peripheral views to the center view.
We follow the basic registration architecture by Genser et al.~\cite{genser_camsi_2020}, which was proven to be effective.
We introduce a cross spectral disparity estimation network which is trained using a standard RGB stereo database by augmenting the RGB in a novel way to generate pseudo spectral data.
This pseudo spectral data generation is presented in Section~\ref{sec:da}.
After warping the peripheral images to the center view, occluded pixels in the warped peripheral views need to be found.
For this, we establish an occlusion detection algorithm that warps the disparity map layer by layer to the peripheral view.
During this process, the pixels that are occluded by foreground objects are marked as missing.
Finally, these pixels need to be reconstructed, for which we introduce a deep guided neural network.
This deep guided neural network leverages the structure of the fully available center view by building local models of the available pixels of the peripheral view and the corresponding pixels of the center view.
These models are then used to reconstruct the occluded pixels.
Again, this network is trained using the proposed pseudo spectral data augmentation of Section~\ref{sec:da}.

Note that a calibration process might be necessary when dealing with real-world data to rectify the images according to their position within the array.
This is particularly necessary to fulfill the epipolar constraint~\cite{ishikawa_epipolar_1998}.
For this, any state-of-the-art calibration procedure can be used for inter camera calibration such as~\cite{genser_camsi_2020} in conjunction with~\cite{ransac_1981}.
There, a homography transformation $\homograhpy$ is estimated based on a checkerboard pattern to warp the peripheral images to the center view to match the checkerboard pattern.
Hence, at this stage, the images are only registered for a single depth plane.
In the following, all images of the multispectral camera array are assumed to have been rectified beforehand.

In subsequent sections, the coordinate system $(\calcoord)$ is used for each channel of the calibrated images $\calibrated_\channel[\calcoord]$, where $\channel$ is the channel or view of the multispectral image.
The destination coordinate system $(\descoord)$ is the one of the center camera $\centercam$, thus $(m_\centercam, n_\centercam) = (\descoord)$.

\subsection{Cross Spectral Disparity Estimation}
\label{subsec:csdl}

The goal of the presented multispectral image registration is to overlay the same pixel of different views on top of each other in the center view.
The distance between the same object pixel in different views is called disparity and is dependent on the depth between the camera array and the object.
This disparity can be estimated from all camera views to the center view.
For this, the views are rotated according to their position such that the disparity between the views is purely horizontal.
For example, to estimate disparities between the camera in the top left and the center camera, both images need to be rotated by 45 degrees clockwise.

To estimate disparity maps, IGEV~\cite{xu_igev_2023} is used, which ranks top across nearly all benchmarks and has a fairly fast execution time.
Furthermore, an iterative update step of an initial disparity estimation is performed, where the number of steps can be controlled for a trade-off between runtime and accuracy.
Unfortunately, this network is trained on RGB images for the left and right view.
In the case of multispectral image registration, left and right views are grayscale images from different spectral bands.
Therefore, the texture of the same object can have completely different grayscale values.
Hence, this network needs to be retrained using spectral data instead of RGB images.
However, there are no multispectral stereo databases for training disparity estimation networks available.
In contrast, there are a lot of RGB stereo datasets available for training.
Therefore, we proposed a novel data augmentation method to generate pseudo spectral images, which is described in Section~\ref{sec:da}.
Using this procedure, always the latest state-of-the-art networks for disparity estimation can be used.
For training, the original training parameters of IGEV are used.

As shown in \fig\ref{fig:disparity}, the disparity estimation starts by rotating the current peripheral view $\calibrated_\channel[\calcoord]$ and the center view $\calibrated_\centercam[\descoord]$ according to the angle of peripheral camera with respect to the center camera such that the disparity is purely horizontal.
Afterwards, a disparity estimation network $\disparityoperator(\cdot)$ is applied on the rotated peripheral view and the correspondingly rotated center view
\begin{equation}
    \disparity^{r}_\channel[\descoord] = \disparityoperator\left( \calibrated^{r}_\centercam[\descoord], \calibrated^{r}_\channel[\calcoord] \right),
\end{equation}
where $\calibrated^{r}_\centercam[\descoord]$ and $\calibrated^{r}_\channel[\calcoord]$ are the rotated images of $\calibrated_\centercam[\descoord]$ and $\calibrated_\channel[\calcoord]$, respectively.
The result is an equivalently rotated disparity estimation $\disparity^{r}_\channel[\descoord]$, which needs to be rotated back to $\disparity_\channel[\descoord]$.
During this conversion the different baselines of the peripheral cameras to the center camera are respected, i.e., the disparity estimates of the diagonal peripheral cameras are divided by a factor of $\sqrt{2}$.
This results in $\cameras - 1$  disparity maps from all the peripheral views to the center view, where $\cameras$ is the number of cameras.
These disparity maps will be of different quality in different regions.
This originates from the fact that different views contain different spectral content.
Since disparity maps for the center view are calculated, all estimated disparity maps $\disparity_\channel[\descoord]$ can be fused to improve the final result.
For this, the median operation is suitable
\begin{equation}
    \disparity[\descoord] = \med_\channel\left( \disparity_\channel[\descoord] \right).
\end{equation}
This disparity map can then be used to map the peripheral views to the center view.

\subsection{Warping and Occlusion Detection}
\label{subsec:warp}

We use the estimated disparity map $\disparity[\descoord]$ to warp the peripheral views to the center view.
Since the disparity map is in the coordinate system of $(\descoord)$, the disparity needs to be added to the destination pixel position to get the position of the peripheral pixel to pull.
This can be formulated as
\begin{equation}
    \begin{pmatrix}
        \descoordX + \rotangle_{\channel, x} \disparity[\descoord]\\
        \descoordY + \rotangle_{\channel, y} \disparity[\descoord]
    \end{pmatrix}
    =
    \begin{pmatrix}
        \calcoordM_\channel\\
        \calcoordN_\channel
    \end{pmatrix},
\end{equation}
where $\rotangle_{\channel, x}$ and $\rotangle_{\channel, y}$ are responsible for including the camera position relative to the center camera.
For purely horizontal aligned cameras, $\rotangle_{\channel, y}$ will be zero and $\rotangle_{\channel, x}$ will contain the relative baseline, for purely vertical cameras vice versa.
For diagonal aligned cameras, $\rotangle_{\channel, x}$ and $\rotangle_{\channel, y}$ will be non-zero and will compensate for the increased baseline.

\begin{figure}
    \centering
    \input{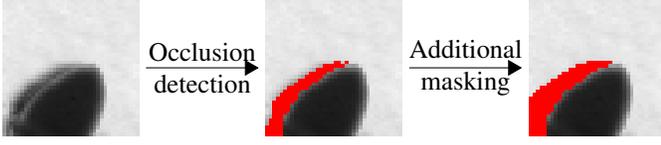}
    \caption{On the left a warped peripheral view is depicted. The darker part is the foreground, which is repeated in the occluded area in the background. In the middle, occluded pixels (red) are found by the proposed detection algorithm. Moreover, one can also see that unsharp pixels cause problems since they belong to foreground and background simultaneously. On the right, these pixels are marked as occluded as well.}
    \label{fig:unsharp}
\end{figure}

The problem with warping the peripheral views using the center disparity is that in occluded regions the occluding objects of the peripheral view will be repeated.
This problem is depicted in the left image of \fig\ref{fig:unsharp}, where the black object in the foreground is repeated in the background to the left.
Thus, we introduce a novel procedure to detect these regions by only using the disparity map.

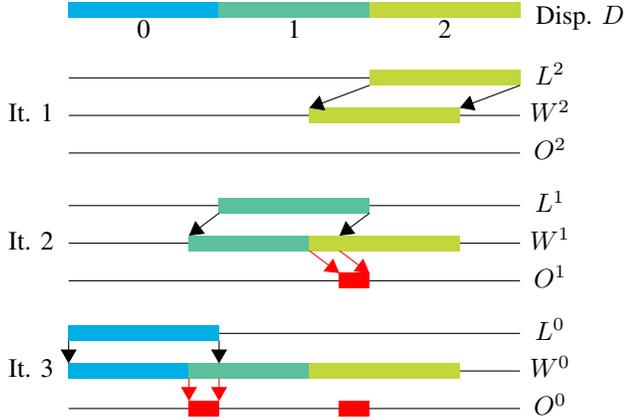
\begin{figure}
    \centering
    \begin{tikzpicture}[y=-1cm, >=triangle 60]
    \fill[cyan!95!lime] (0, 0) rectangle ++(2,0.2);
    \fill[cyan!50!lime] (2, 0) rectangle ++(2,0.2);
    \fill[cyan!5!lime] (4, 0) rectangle ++(2,0.2);

    \node[] at (1, 0.35) {0};
    \node[] at (3, 0.35) {1};
    \node[] at (5, 0.35) {2};

    \node[] at (6.8, 0.2) {Disp. $\disparity$\strut};

    \begin{scope} [shift={(0, 1)}]
        \draw [-] (0, 0) -- (6, 0);
        \fill[cyan!5!lime] (4, -0.1) rectangle ++(2,0.2);
        \draw [-] (0, 0.5) -- (6, 0.5);
        \fill[cyan!5!lime] (3.2, 0.4) rectangle ++(2,0.2);
        \node[] at (6.4, 0) {$\disparitylayer^2$\strut};
        \node[] at (6.4, 0.5) {$\disparitywarped^2$\strut};

        \draw [->] (4, 0.1) -- (3.2, 0.4);
        \draw [->] (6, 0.1) -- (5.2, 0.4);

        \node[] at (6.4, 1) {$\mask^2$\strut};
        \draw [-] (0, 1) -- (6, 1);

        \node[] at (-0.5, 0.5) {It. 1\strut};
    \end{scope}

    \begin{scope} [shift={(0, 2.7)}]
        \draw [-] (0, 0) |- (6, 0);
        \fill[cyan!50!lime] (2, -0.1) rectangle ++(2,0.2);
        \draw [-] (0, 0.5) |- (6, 0.5);
        \fill[cyan!50!lime] (1.6, 0.4) rectangle ++(2,0.2);
        \fill[cyan!5!lime] (3.2, 0.4) rectangle ++(2,0.2);
        \node[] at (6.4, 0) {$\disparitylayer^1$\strut};
        \node[] at (6.4, 0.5) {$\disparitywarped^1$\strut};

        \draw [->] (2, 0.1) -- (1.6, 0.4);
        \draw [->] (4, 0.1) -- (3.6, 0.4);

        \node[] at (6.4, 1) {$\mask^1$\strut};
        \draw [-] (0, 1) -- (6, 1);
        \fill[red] (3.6, 0.9) rectangle ++(0.4,0.2);

        \draw [->, red] (3.2, 0.6) -- (3.6, 0.9);
        \draw [->, red] (3.6, 0.6) -- (4, 0.9);

        \node[] at (-0.5, 0.5) {It. 2\strut};
    \end{scope}

    \begin{scope} [shift={(0, 4.4)}]
        \draw [-] (0, 0) |- (6, 0);
        \fill[cyan!95!lime] (0, -0.1) rectangle ++(2,0.2);
        \draw [-] (0, 0.5) |- (6, 0.5);
        \fill[cyan!95!lime] (0, 0.4) rectangle ++(2,0.2);
        \fill[cyan!50!lime] (1.6, 0.4) rectangle ++(2,0.2);
        \fill[cyan!5!lime] (3.2, 0.4) rectangle ++(2,0.2);
        \node[] at (6.4, 0) {$\disparitylayer^0$\strut};
        \node[] at (6.4, 0.5) {$\disparitywarped^0$\strut};

        \draw [->] (0, 0.1) -- (0, 0.4);
        \draw [->] (2, 0.1) -- (2, 0.4);

        \node[] at (6.4, 1) {$\mask^0$\strut};
        \draw [-] (0, 1) -- (6, 1);
        \fill[red] (3.6, 0.9) rectangle ++(0.4,0.2);
        \fill[red] (1.6, 0.9) rectangle ++(0.4,0.2);

        \draw [->, red] (1.6, 0.6) -- (1.6, 0.9);
        \draw [->, red] (2, 0.6) -- (2, 0.9);

        \node[] at (-0.5, 0.5) {It. 3\strut};
    \end{scope}
\end{tikzpicture}
    \caption{The process of finding occlusions for a single depth map line. Occluded pixels are shown in red.}
    \label{fig:occlusion}
\end{figure}

This novel algorithm, of which three iterations are shown in \fig\ref{fig:occlusion}, starts by warping the disparity map layer by layer to the peripheral view starting from the layer with the highest disparity, since a higher disparity indicates an object being in the foreground, which occludes background objects with a lower disparity.
The disparity layers for integer layer $d \in [\lfloor\min(\disparity[\descoord])\rfloor, \lceil\max(\disparity[\descoord])\rceil]$ are given by
\begin{gather}
    \disparitylayer^d[\descoord] =
    \begin{cases}
        \disparity[\descoord],   & \text{if } \disparity[\descoord] - \layeroverlap \leq d \leq \disparity[\descoord] + \layeroverlap\\
        0,                  & \text{else}
    \end{cases},
\raisetag{10pt}
\end{gather}
where $\layeroverlap \geq 0.5$ is a constant controlling the amount of overlap between different disparity layers.
An overlap is necessary to avoid integer rounding issues for the following steps.
After extraction, the disparity layers $\disparitylayer^d[\descoord]$ are warped according to their disparity index to the peripheral view
\begin{equation}
    \disparitylayer^d_w[\descoord] = \disparitylayer^d[\descoordX + d, \descoordY].
\end{equation}
Then, as shown in \fig\ref{fig:occlusion}, the occlusion detection iterates over the warped disparity layers $\disparitylayer^d_w[\descoord]$ starting from the highest disparity.
For each iteration, the cumulated warped disparity $\disparitywarped^{d}[\descoord]$, an all-zeros image in the beginning, is being kept track of
\begin{gather}
    \disparitywarped^{d}[\descoord] =
    \begin{cases}
        \disparitylayer^d_w[\descoord],   & \text{if } \disparitylayer^d_w[\descoord] \geq \disparitywarped^{d + 1}[\descoord]\\
        \disparitywarped^{d + 1}[\descoord], & \text{else}
    \end{cases}.
\raisetag{10pt}
\end{gather}
Furthermore, using this warped disparity, the occlusion mask viewed from the center can be iteratively updated
\begin{gather}
\begin{split}
    &\mask^{d}[\descoordX - d, \descoordY] = \\
    &=\begin{cases}
        1   & \text{if } \disparitylayer^d_w[\descoord] \leq \disparitywarped^{d + 1}[\descoord] - \phi\\
        \mask^{d + 1}[\descoordX - d, \descoordY], & \text{else}
    \end{cases},
\end{split}
\raisetag{10pt}
\end{gather}
where $\phi$ is the disparity distance needed that a foreground pixel occludes a background pixel.
While $\phi$ is set to $0.5$, the aforementioned $\layeroverlap$ is chosen to be $0.75$, which were found using the training set of SceneFlow~\cite{sceneflow_2016}.
The final mask is $\mask[\descoord] = \mask^{\lfloor\min(\disparity[\descoord])\rfloor}[\descoord]$, where $\mask^{\lfloor\min(\disparity[\descoord])\rfloor}[\descoord]$ is the occlusion map of the final iteration and thus contains all occlusions.
This procedure has to be performed for all peripheral views individually according to their rotation with respect to the center view, which yields masks $\mask_\channel[\descoord]$ for all peripheral views.

Unfortunately, unsharp pixels, caused by motion blur or depth of field blur, contain content from background and foreground, but are assigned to a single disparity value.
This problem is depicted in the center image of \fig\ref{fig:unsharp}, where there are still some darker pixels in the background left.
Thus, unsharp pixels that belong to the background need to be masked as well.
Otherwise, they would negatively influence the reconstruction process by yielding wrong grayscale values directly next to missing pixel areas.
To cope with this behaviour, $\addmask$ additional pixels in the direction of masking next to already occluded pixels are marked as missing as well, which is shown in the right image of \fig\ref{fig:unsharp}.
This can be achieved by adding translated versions of the original mask.
Of course, all pixels need to be clipped between zero and one, again.
This technique can be viewed as a direction-dependent dilation of occluded areas.
The parameter $\addmask$ needs to be adapted on the sharpness of the image in all layers.

\subsection{Cross Spectral Reconstruction}

\begin{figure*}
    \centering
    \input{images/tikz/rec_arch.tikz}
    \caption{Architecture of our deep guided reconstruction network. The blue blocks indicate convolutional layers using classical convolutions with parameters channels/kernel size/stride.}
    \label{fig:rec_architecture}
\end{figure*}

The last step within the registration is to reconstruct the occluded pixels.
Fortunately, the image of the center camera is fully available and can serve as guide.
However, the center camera records in a different spectral area in comparison to the peripheral cameras.
Therefore, the goal is to exploit the structure of the guide to reconstruct the missing pixels of the peripheral images.

An overview over the proposed architecture of the deep guided neural network is given in \fig\ref{fig:rec_architecture}.
The goal of this whole network is to estimate linear regression coefficients applied to the fully available center view $\msmasked_\centercam[\descoord]$ to predict the missing pixels.
The input images for the network are the reference image $\msmasked_\centercam[\descoord]$, i.e., the center view, the occluded image for the view to reconstruct $\msmasked_\channel[\descoord]$ and the corresponding mask $\mask_\channel[\descoord]$.
These images are concatenated and fed into a CNN with four downsampling layers with stride two, which is responsible for predicting two low resolution 3D cubes of linear regression coefficients.
The number of downsampling layers determines how many high resolution pixels are falling into one bin of the resulting two 3D cubes, one for each linear regression parameter.
In our case, four downsampling layers lead to a bin size of $16 \times 16$.
Afterwards, a couple of convolutional layers are applied to increase the field of view to find the appropriate pixels to set up the linear regression parameters.
Finally, the last convolutional layer is responsible for creating the proper number of channels such that the output can be transformed into the 3D cubes of linear regression coefficients.
Since 32 luma bins are chosen and each bin needs two coefficients for linear regression, the resulting channel size is 64.

This low resolution cube serves as interpolated lookup table which is introduced in bilateral guided upsampling~\cite{bilateral_chen_2016, gharbi_deep_2017}.
The idea is that different grayscale values correspond to different objects and textures, and thus need different linear regression coefficients.
Thus, even though this cube has a lower spatial resolution, edges can be well reconstructed assuming different grayscale values for different objects.
The granularity of this behaviour is set by the number of luma bins.

To slice out the high resolution linear regression image $\highLRA_{\channel}[\descoord]$ from the low resolution cube $\lowLRA_{\channel}[i, j, k, l]$, the interpolation~\cite{sippel_deepguided_2023}
\begin{gather}
    \highLRA_{\channel}[\descoord]\hspace{-0.1cm}=\hspace{-0.3cm}\sum_{i, j, k, l}\hspace{-0.1cm}\interpolationkernel( \scalex \descoordX\hspace{-0.1cm}-\hspace{-0.1cm}i) \interpolationkernel( \scaley \descoordY\hspace{-0.1cm}-\hspace{-0.1cm}j) \interpolationkernel( \scaler \msmasked_\centercam[\descoord]\hspace{-0.1cm}-\hspace{-0.1cm}k) \lowLRA_{\channel}[i, j, k, l]
\raisetag{10pt}
\end{gather}
is used, where $\interpolationkernel(\cdot) = \max(1 - |\cdot|, 0)$ is the linear interpolation kernel, $\scalex$ and $\scaley$ are the ratios of the spatial dimensions of the 3D cubes with respect to the high resolution.
Accordingly, $\scaler$ is the ratio of luma bins to the maximum intensity value.
In the exact same way, the high resolution linear regression bias image $\highLRB_{\channel}[\descoord]$ is extracted from the low resolution bias cube $\lowLRB_{\channel}[i, j, k, l]$.
Finally, the network output can be calculated by applying the linear regression coefficients
\begin{equation}
	\networkoutput_{\channel}[\descoord] = \highLRA_{\channel}[\descoord] \cdot \msmasked_\centercam[\descoord] + \highLRB_{\channel}[\descoord].
\end{equation}
Of course, the final result will only use the network output for the occluded pixels
\begin{equation}
	\msimage_{\channel}[\descoord] = (1 - \mask_{\channel}[\descoord]) \cdot \msmasked_{\channel}[\descoord] + \mask_{\channel}[\descoord] \cdot \networkoutput_{\channel}[\descoord].
\end{equation}

The network can work with any image size as only convolutional layers are used.
However, the images need to be padded, so that the bins cover the image perfectly.
The $l_1$-loss in occluded areas as well as the multiscale structural similarity index~\cite{ms_ssim_2003} between the network output $\networkoutput_{\channel}[\descoord]$ and the ground-truth images $\hat{\msimage}_{\channel}[\descoord]$
\begin{multline}
	L(\networkoutput_{\channel}[\descoord], \hat{\msimage}_{\channel}[\descoord], \mask_{\channel}[\descoord]) = \\
    \frac{1 - \lossalpha}{\text{sum}(\mask_\channel[\descoord])}
    \sum_{\descoord}
    \begin{cases}
        |\networkoutput_{\channel}[\descoord] - \hat{\msimage}_{\channel}[\descoord]|,& \text{if} \ \mask_\channel[\descoord] = 1 \\
    0 & \text{else},
	\end{cases} \\
    + \lossalpha \cdot (1 - \text{MS-SSIM}(\networkoutput_{\channel}[\descoord], \hat{\msimage}_{\channel}[\descoord]))
\raisetag{10pt}
\end{multline}
are used to train the network, where $\text{MS-SSIM}(\cdot)$ calculates the multiscale structural similarity index.
The other component of the loss function is the $l_1$-loss of the occluded pixels.
Therefore, the $l_1$-loss ignores the known pixels, which are already part of the MS-SSIM loss.
The weight factor $\lossalpha$ is set to 0.84 as proposed by Zhao et al.~\cite{loss_alpha_2017}.

\begin{figure}
    \centering
    \input{images/tikz/rec_db.tikz}
    \caption{The generation process of the adapted database using SceneFlow as base. Since the proposed cross spectral disparity estimation, warping and mask detection procedure are used, the network can learn to adapt to the resulting artefacts. The pseudo spectral data augmentation is executed during training.}
    \label{fig:rec_db}
\end{figure}

The training is done on an adapted dataset based on SceneFlow~\cite{sceneflow_2016}.
Realistic masks can be generated, when directly using the proposed methods from the previous steps in the pipeline, i.e., the proposed cross spectral disparity estimation and the proposed warping and occlusion detection.
This adaptation process is shown in \fig\ref{fig:rec_db} and described in the following.
The base for the generated database are the left and right images of SceneFlow~\cite{sceneflow_2016}.
The red channel of the left image and the blue channel of the right image is used to estimate a disparity map in a cross spectral way.
The reason behind this is that the red component and the blue component of an RGB image are furthest away in a spectral sense.
Therefore, the disparity estimator should be worse than any other combination and thus produces most artefacts.
The cross spectral disparity estimator of Section~\ref{subsec:csdl} is used for that.

Afterwards, this estimated disparity is used to warp the right RGB image of the SceneFlow database to the left perspective.
Subsequently, the occlusion detection method then also produces an occlusion map for the estimated disparity.
The warping and occlusion detection from Section~\ref{subsec:warp} is applied.

As shown in \fig\ref{fig:rec_db}, the proposed database then contains the original left RGB image of SceneFlow used as guide image and the warped RGB image, which represents the peripheral view.
Additionally, this database contains the occlusion maps, which state which pixels of the peripheral view need to be reconstructed.
For training, both RGB images are transformed to pseudo spectral grayscale images using the data augmentation presented in Section~\ref{sec:da}.
Additionally, random flips and axes swaps are added to simulate four warping directions.
Apart from these flips, the occlusion map stays as it is.

The reconstruction network is trained for 1024 epochs on this custom database.
The learning rate is set to 0.0001 and is halved at epochs 256, 512 and 756.
Adam is used as optimizer with parameters $\beta_1=0.5$ and $\beta_2=0.999$.

\section{Data Augmentation}
\label{sec:da}
\begin{figure}
    \centering
    \input{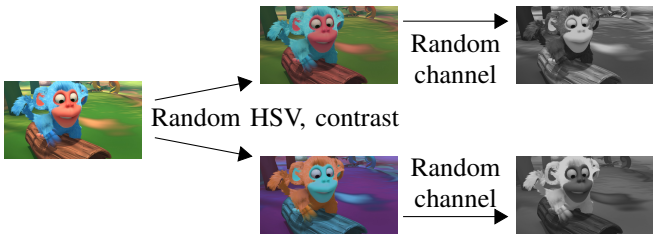}
    \vspace*{-0.2cm}
    \caption{Two example of the procedure to create pseudo spectral image from an RGB image.}
    \label{fig:da_spectral}
    \vspace*{-0.3cm}
\end{figure}

Pseudo spectral data augmentation plays a crucial role in training the cross spectral disparity estimation as well as the cross spectral reconstruction.
The problem in training neural networks for multispectral imaging lies in the limited availability of multispectral training databases.
In contrast, there are several datasets with millions of RGB images ready to train on.
Moreover, no databases exist for cross spectral disparity estimation, while stereo RGB datasets are not rare.
Hence, the goal is to leverage this data availability for training multispectrally.
To achieve this, a fast spectral image generation from RGB images is necessary.

The idea is to change the grayscale values of different objects, while the structure of the image should be preserved.
Since only wideband red, green and blue components are provided, the generated spectral images will not match real images from different spectral bands perfectly.
Thus, it is necessary to generate diverse augmented spectral images.
Hence, this whole process is called pseudo spectral image generation.

The process of generating pseudo spectral grayscale images is depicted in \fig\ref{fig:da_spectral}, where two examples are generated.
The first step is to alternate the spectral response of the RGB image, which can be easily changed by jittering the hue randomly.
This is achieved by rotating the HSV cone according to a random factor with a uniform distribution.
Similarly, brightness, saturation and contrast of the input RGB images can also be adapted according to random factors with a uniform distribution.
While the hue can be fully rotated to each possible angle, brightness, saturation and contrast are individually chosen from a uniform distribution in the range $[0.5, 1.5]$.
Hence, overexposed environments are simulated as well.
Moreover, the permutation of applying these four augmentations changes to the RGB image is randomly chosen as well.
In the final step, a random channel of the resulting RGB image is selected as single spectral band image.
The resulting grayscale images can then be used to train the cross spectral disparity estimation network as well as the deep guided neural network.

\section{Evaluation}
\label{sec:evaluation}
As main evaluation database, we choose the synthetic hyperspectral video array database (HyViD)~\cite{sippel_synthetic_2023}.
This database contains 7 scenes rendered for 30 frames from a camera array very similar to the one in \fig\ref{fig:task}.
Furthermore, 31 hyperspectral channels were rendered for each camera.
Thus, synthetic filters can be simulated on this data.
For all upcoming evaluations using HyViD, we simulated bandpass filters with varying bandwidths at center wavelengths 425 nm, 450 nm, 470 nm, 500 nm, 525 nm, 550 nm, 600 nm, 633 nm and 650 nm.
The resolution of the resulting multispectral images is 1600 \texttimes\ 1200.
Note that the cameras are perfectly aligned in this synthetic setup.
Hence, no calibration procedure is required.

\subsection{Data Augmentation}
\begin{table}[t]
	\small
	\centering
	\caption{Evaluation of the proposed data augmentation in comparison to the pseudo spectral image generation of~\cite{sippel_deepguided_2023} for the proposed reconstruction network. The evaluation is based on HyViD~\cite{sippel_synthetic_2023}. Higher values are better.}
	\label{tab:da}
	\begin{tabular}{lcc}
				& PSNR & SSIM \\
		\hline
				\cite{sippel_deepguided_2023}   & 39.19             & 0.971 \\

				Proposed                        & \textbf{39.40}    & \textbf{0.973} \\
	\end{tabular}
\end{table}

First, we evaluate the proposed data augmentation in comparison to our previous pseudo spectral data generation from~\cite{sippel_deepguided_2023}.
For that, the proposed deep guided reconstruction network was trained using both data augmentations.
The evaluation is based on the full registration task on HyViD, while only the reconstruction network is exchanged.
As shown in \tab\ref{tab:da}, the proposed data augmentation is slightly superior in terms of Peak Signal-to-Noise Ratio (PSNR) and Structural Similarity Index (SSIM).
While the gain of about 0.2 dB of the proposed data augmentation in comparison to a previous version is relatively small, it comes for free in terms of runtime during inference.

\subsection{Disparity Estimation}
\begin{table}[t]
	\small
	\centering
	\caption{Evaluation of the cross spectral disparity estimation on HyViD~\cite{sippel_synthetic_2023}. The results are given by PSNR in dB / SSIM.}
	\label{tab:eval_depth}
	\begin{tabular}{@{\hspace*{0.0cm}}lcccc@{\hspace*{0.0cm}}}
        & Genser & CSDL & CADE & Proposed \\
        & \cite{genser_camsi_2020} & \cite{genser_deep_2020} & \cite{sippel_cade_2024} & IGEV~\cite{xu_igev_2023} \\
        \hline
        Fam. hou.           & 34.44/.967 & 32.52/.950 & 33.92/.962 & \textbf{36.11/.976} \\
        Med. sea.           & 43.78/.990 & 44.20/.990 & 45.85/.992 & \textbf{46.89/.993} \\
        City                & 44.01/.991 & 42.66/.991 & 45.40/.993 & \textbf{46.78/.995} \\
        Outdoor             & 33.01/.940 & 32.40/.923 & 32.96/.932 & \textbf{33.89/.943} \\
        Forest              & 30.37/.916 & 28.64/.863 & 30.54/.906 & \textbf{31.40/.920} \\
        Indoor              & 36.85/.983 & 43.11/.992 & \textbf{44.80/.993} & 42.24/.992 \\
        Lab                 & 31.50/.969 & 37.35/.987 & \textbf{39.08/.991} & 38.48/.990 \\
        \hline
        Average             & 36.28/.965 & 37.27/.957 & 38.94/.967 & \textbf{39.40/.973} \\
	\end{tabular}
\end{table}

\begin{table}[t]
	\small
	\centering
	\caption{Evaluation of different occlusion detection algorithms suitable for the presented registration on HyViD~\cite{sippel_synthetic_2023}. The results are given by PSNR in dB / SSIM.}
	\label{tab:eval_mask}
	\begin{tabular}{@{\hspace*{0.0cm}}lcccc@{\hspace*{0.0cm}}}
        & SymmNet & SymmNet-D & Genser & Proposed \\
        & \cite{li_symmnet_2018} & \cite{li_symmnet_2018} & \cite{genser_camsi_2020} & \\
        \hline
        Fam. hou.           & 31.56/.952 & 35.80/\textbf{.981} & 34.03/.961 & \textbf{36.11}/.976 \\
        Med. sea.           & 43.40/\textbf{.993} & 43.59/\textbf{.993} & 46.80/\textbf{.993} & \textbf{46.89/.993} \\
        City                & 41.19/.992 & 42.44/.994 & \textbf{47.38/.995} & 46.78/\textbf{.995} \\
        Outdoor             & 31.27/.933 & 33.19/\textbf{.953} & 33.05/.932 & \textbf{33.89}/.943 \\
        Forest              & 26.71/.856 & 29.82/.916 & 28.90/.872 & \textbf{31.40/.920} \\
        Indoor              & 34.95/.981 & 36.57/.987 & 41.44/.990 & \textbf{42.24/.992} \\
        Lab                 & 32.48/.977 & 33.82/.984 & \textbf{39.03}/.988 & 38.47/\textbf{.990} \\
        \hline
        Average             & 34.51/.955 & 36.46/.972 & 38.66/.962 & \textbf{39.40/.973} \\
	\end{tabular}
\end{table}

We evaluated the cross spectral disparity estimation methods within the registration pipeline using the ground truth registered images of HyViD.
For that, the proposed occlusion detection and reconstruction network are used after the disparity estimation.
The method by Genser~\cite{genser_camsi_2020} is a non-learning approach based on calculating the zero normalized cross correlation between different windows.
Cross Spectral training for Deep Learning (CSDL)~\cite{genser_deep_2020} retrains PSMNet~\cite{chang_psmnet_2018} using a simple spectral augmentation procedure.
In our previous publication, CADE~\cite{sippel_cade_2024} retrains GANet~\cite{zhang_ganet_2019} using a slightly improved data augmentation and local normalized images.
As \tab\ref{tab:eval_depth} shows, the proposed method of retraining IGEV~\cite{xu_igev_2023} using the proposed pseudo spectral augmentation outperforms all other estimators in terms of PSNR and SSIM on average.
Note that GANet of CADE~\cite{sippel_cade_2024} performs better for scenes \textit{indoor} and \textit{lab}, which contains a lot less structured regions where IGEV performs worse than the guided aggregation approach.
It is also noteworthy that the GANet-based approach takes approximately 45 seconds for each multispectral image, while IGEV only requires less than 5 seconds as shown in \tab\ref{tab:eval_runtime}.

\subsection{Occlusion Detection}

We compare our proposed occlusion detection against the occlusion detection by Genser et al.~\cite{genser_camsi_2020} and SymmNet~\cite{li_symmnet_2018}.
SymmNet is a neural network predicting the occlusion from left and right view images.
Therefore, this network has been retrained using the proposed spectral image augmentation.
Furthermore, since the other methods work on the disparity map and not on the left and right images, the network has been modified and retrained to only take the disparity map as input (SymmNet-D).
Again, the evaluation is based on HyViD.
Thus, this time, the occlusion detection method is exchanged, while the disparity estimator and reconstruction network stays as proposed.
As \tab\ref{tab:eval_mask} shows, SymmNet-D exploiting the disparity map performs slightly better than its original version.
Our proposed method outperforms all other methods in terms of PSNR and SSIM for almost every scene.
On average, the novel occlusion detection outpeforms the method by Genser et al.~\cite{genser_camsi_2020} by more than 0.7 dB in terms of PSNR as well as by a significant margin in terms of SSIM.

\subsection{Reconstruction}

\begin{table*}[t]
	\small
	\centering
	\caption{Average PSNR in dB and SSIM of different reconstruction methods on HyViD~\cite{sippel_synthetic_2023}. The results are given by PSNR in dB / SSIM.}
	\label{tab:eval_csr}
	\begin{tabular}{@{\hspace*{0.0cm}}lcccccc@{\hspace*{0.0cm}}}
		&	MAT	& FSR & NOCS & Genser & DGNet & Proposed \\
        & \cite{li_mat_2022} & \cite{genser_spectral_2018} & \cite{sippel_spatio_2021} & \cite{genser_camsi_2020} & \cite{sippel_deepguided_2023} & \\
        \hline
        Family house        & 27.66/.911 & 28.53/.917 & 31.22/.935 & 34.41/.969 & 35.87/.975 & \textbf{36.11/.976} \\
        Medieval seaport    & 40.95/.989 & 39.55/.989 & 42.41/.990 & 44.70/.992 & 46.67/\textbf{.993} & \textbf{46.89/.993} \\
        City                & 39.47/.990 & 38.10/.989 & 41.13/.991 & 43.24/.994 & 46.68/\textbf{.995} & \textbf{46.78/.995} \\
        Outdoor             & 27.77/.870 & 27.63/.878 & 30.10/.899 & 32.11/.930 & 33.74/.942 & \textbf{33.89/.943} \\
        Forest              & 22.95/.735 & 23.45/.762 & 25.53/.796 & 27.61/.871 & 30.55/.909 & \textbf{31.40/.920} \\
        Indoor              & 33.98/.975 & 32.26/.973 & 36.64/.981 & 36.34/.986 & 41.68/.991 & \textbf{42.24/.992} \\
        Lab                 & 30.14/.958 & 31.09/.970 & 33.10/.974 & 34.03/.984 & 38.02/.989 & \textbf{38.48/.990} \\
        \hline
        Average             & 31.85/.918 & 31.52/.926 & 34.30/.938 & 36.06/.961 & 39.03/.970 & \textbf{39.40/.973} \\
	\end{tabular}
\end{table*}

Different reconstruction algorithms are evaluated on HyViD.
For that, the proposed disparity estimation is used.
Afterwards, the proposed occlusion detection algorithm detects the pixels to reconstruct.
The inpainting methods MAT~\cite{li_mat_2022}, which is a transformer, and the frequency selective reconstruction (FSR)~\cite{genser_spectral_2018} are compared against the guided reconstruction methods NOCS~\cite{sippel_spatio_2021}, the reconstruction algorithm from Genser et al.~\cite{genser_camsi_2020}, our previous reconstruction network DGNet~\cite{sippel_deepguided_2023} and our proposed reconstruction network.
Note that the guided reconstruction methods significantly outperform the inpainting methods by exploiting the structure of the center view.
\tab\ref{tab:eval_csr} shows that the proposed method outperforms all other methods in terms of PSNR and SSIM and also beats our previous network by nearly 0.4 dB in terms of PSNR.
Note that, the proposed network improves PSNR on all scenes while gaining most on the scenes \textit{forest}, \textit{indoor} and \textit{lab}.
SSIM is also improved for most scenes and kept at the same level for the other two scenes, which already have a very good reconstruction quality.
In comparison to the reconstruction method by Genser et al.~\cite{genser_camsi_2020}, the proposed reconstruction network improves the reconstruction in the proposed pipeline by more than 3 dB in terms of PSNR as well as in terms of SSIM.

\subsection{Complete Registration Pipeline}

\begin{figure*}
    \centering
    \input{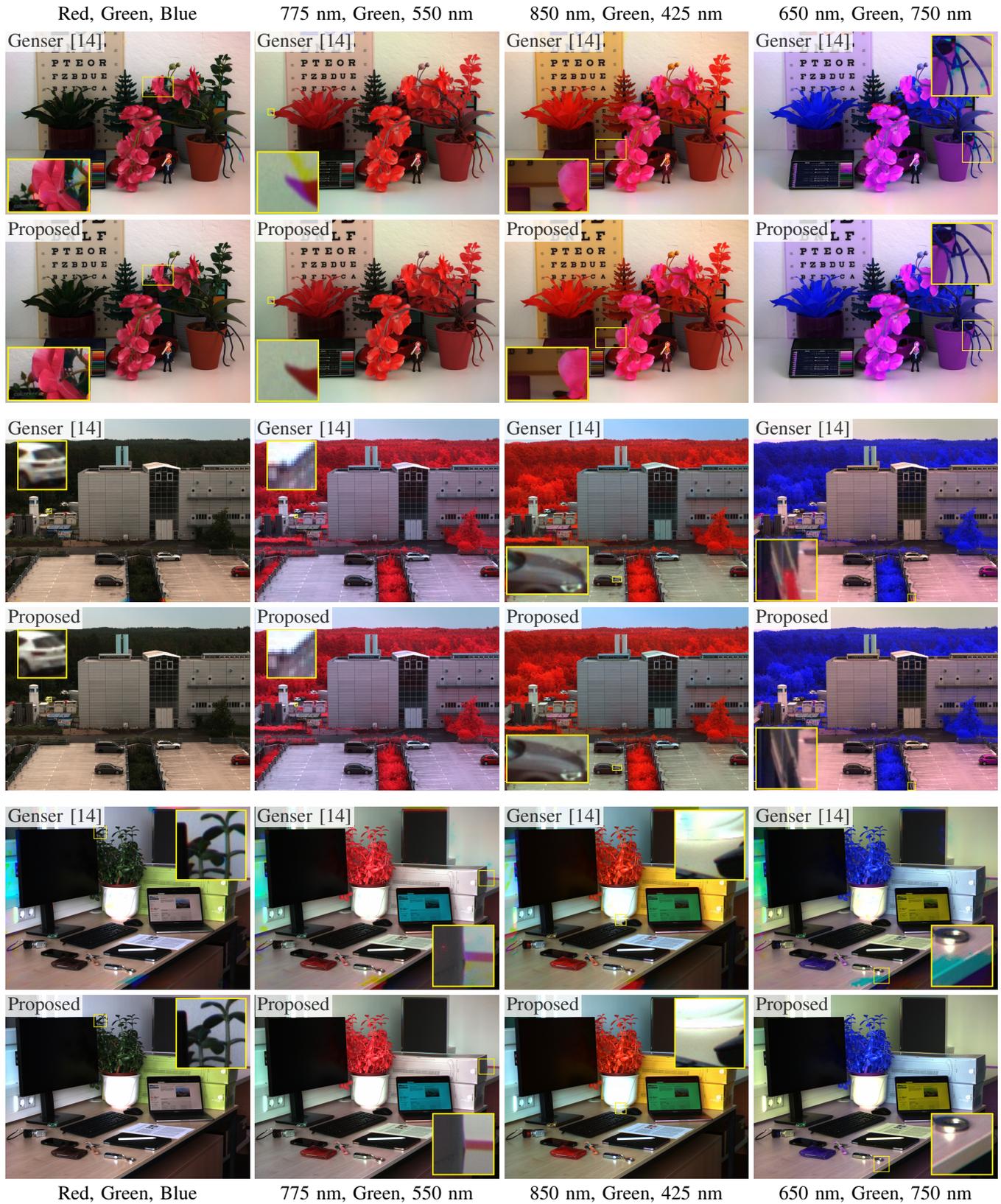}
	\caption{Registration results of three real-world records of CAMSI. On the left of each scene, an RGB image is composed using the center row of the camera array (cameras 4, 5, 6). The other images are false color images, where the red and blue channel show different channels of the array, while the green channel always contains the center camera, where a green bandpass filter is mounted in front of the camera. Digital version along with zooming in recommended.}
	\label{fig:real_data}
\end{figure*}
\begin{table}[t]
	\small
	\centering
	\caption{Evaluation of the complete registration against the procedure proposed by Genser et al.~\cite{genser_camsi_2020}. The results are given by PSNR in dB / SSIM.}
	\label{tab:eval_registration}
	\begin{tabular}{@{\hspace*{0.0cm}}l@{\hspace*{0.1cm}}ccc@{\hspace*{0.0cm}}}
        & Genser~\cite{genser_camsi_2020} & Proposed \\
        \hline
        Family house        & 32.03/.946 & \textbf{36.11/.976} \\
        Medieval seaport    & 42.68/.987 & \textbf{46.89/.993} \\
        City                & 43.33/.990 & \textbf{46.78/.995} \\
        Outdoor             & 31.62/.916 & \textbf{33.89/.943} \\
        Forest              & 26.82/.841 & \textbf{31.40/.920} \\
        Indoor              & 42.10/.990 & \textbf{42.24/.992} \\
        Lab                 & 35.44/.980 & \textbf{38.48/.990} \\
        \hline
        Average             & 36.29/.950 & \textbf{39.40/.973} \\
	\end{tabular}
\end{table}

The only full competitor for our proposed multispectral snapshot image registration is described by Genser et al.~\cite{genser_camsi_2020}.
The PSNR and SSIM values for every scene in HyViD are summarized in \tab\ref{tab:eval_registration}.
The proposed multispectral image registration outperforms~\cite{genser_camsi_2020} in every scene and achieves a total improvement of over 3 dB.

Additionally, \tab\ref{tab:eval_runtime} shows the different runtimes necessary to register and reconstruct CAMSI images for the first frame of scene \textit{family house}.
On a CPU (Intel i9-7940X), the method of Genser et al.~\cite{genser_camsi_2020} needs more than 3 times longer.
As soon as our proposed registration is executed on a GPU (Nvidia RTX 3090), this factor increases to 111.
From \tab\ref{tab:eval_runtime} it gets apparent that the execution time of the disparity estimation needs to be improved if a further reduction in runtime is desired.

\subsection{Real World Images}

\begin{table}[t]
	\small
	\centering
	\caption{Runtime of the different registration methods in seconds. The novel multispectral image registration can be executed on the GPU.}
	\label{tab:eval_runtime}
	\begin{tabular}{@{\hspace*{0.0cm}}l@{\hspace*{0.2cm}}c@{\hspace*{0.2cm}}c@{\hspace*{0.2cm}}c@{\hspace*{0.2cm}}c@{\hspace*{0.0cm}}}
        & Genser~\cite{genser_camsi_2020} & Proposed (CPU) & Proposed (GPU) \\
        \hline
        Disparity       & 57.96 & 166.9 & 4.782 \\
        Occlusion       & 330.7 & 0.946 & 0.082 \\
        Reconstruction  & 161.3 & 1.517 & 0.059 \\
        \hline
        Total   & 549.9 & 169.4 & 4.924 \\
	\end{tabular}
\end{table}

So far, the evaluation was based on synthetic hyperspectral camera array data, since a ground truth is available and objective quantitative values can be obtained.
This evaluation does not reveal whether the proposed multispectral image registration also works for real-world camera array data.
Unfortunately, for this data no ground-truth can be obtained.
Hence, the performance is proven visually.
For that, three scenes using the camera array for multispectral imaging from \fig\ref{fig:task} were recorded and registered.
Note that a calibration method for inter-camera alignment is necessary for these real-world records.
For that, the calibration procedure of Genser et al.\cite{genser_camsi_2020} is used for both methods.
Thus, the influence of calibration is the same for both methods.
Due to the misalignment of the cameras, the borders to reconstruct are much larger than in the case of synthetic data.
RGB and false color images of a lab scene, an outdoor scene and an office scene are shown in \fig\ref{fig:real_data}.
Apart from obvious problems at the borders by the method of Genser~\cite{genser_camsi_2020}, the scenes show the different strengths of the proposed registration.
The lab scene reveals the superiority in high frequency regions.
The outdoor scene shows that the subpixel accuracy of the disparity estimation of the novel multispectral image registration is better, since this scene only contains small disparity changes.
Furthermore, an example of the improved reconstruction at the border of the images is highlighted.
The office scene contains all of the aforementioned advantages.

\section{Conclusion}
\label{sec:conclusion}
In this paper, we introduced an image registration procedure for snapshot multispectral imaging using off-the-shelf hardware.
The multispectral image registration performs three steps, for which novel components have been presented.
First, a cross spectral disparity estimation network was introduced, which is trained on a popular RGB stereo database using the proposed pseudo spectral data augmentation.
Then, we established an occlusion detection algorithm which works on this estimated disparity map.
Finally, a deep guided neural network reconstructs the missing pixels by exploiting the structure of the center camera.
This network was also trained using the proposed pseudo spectral data augmentation.
The evaluation revealed that each single element except calibration is able to outperform all corresponding competitors.
Even more important, the novel registration outperforms the state of the art by over 3 dB in terms of PSNR.
Besides, our proposed registration can be easily executed on a GPU, which significantly improves execution time by a factor of 111.

Our proposed multispectral image registration is able to cover diverse scenarios in a very satisfying manner.
However, this approach on multispectral imaging by using camera arrays has some limitations when the distance between the camera and the scene is too small.
First, specular reflections cannot be properly registered and reconstructed by a multi-camera approach, since the positions of these depend on the viewing angle on the object.
Thus, for different cameras, specular reflections are at different positions on the object in the scene.
Moreover, if the distance between scene and camera array is getting smaller, the disparity increases.
At some point, the disparity exceeds the frame of the image sensors.
This leads to a wrong warping and occlusion detection.
Note that this issue is an inherent problem of any camera array.

In the future, the registration shall be converted to an end-to-end neural network.
Either all the elements and its connections are being made derivable.
Here, the problem of the current pipeline is that the occlusion detection provides a binary classification, whether a pixel is occluded or not.
This binary decision within the pipeline destroys gradients.
Hence, these decision cannot be backpropagated to the disparity estimation network.
Another option is to develop completely new concepts for a multispectral end-to-end registration network.
For example, works in multiplane images~\cite{flynn_mpi_2019, srinivasan_mpi_2019, tucker_mpi_2020} or neural radiance fields~\cite{yu_pixelnerf_2021} can be used as basis for developing an end-to-end registration network.
Moreover, future registration methods could also leverage the temporal axis to further increase quality.

\bibliographystyle{IEEEtran}
\bibliography{refs}

\vfill

\end{document}